\documentstyle[11pt,aaspp4]{article}
\lefthead{Freytag \& Salaris}
\righthead{Stellar Envelope Convection of Globular Cluster Isochrones from
           Hydro-Simulations}

\begin{document}

\title{Stellar Envelope Convection
       calibrated by Radiation Hydrodynamics Simulations:
       Influence on Globular Cluster Isochrones}

\author{Bernd Freytag}
\affil{Astronomical Observatory,
       Juliane Maries Vej 30,
       DK-2100 Copenhagen East,
       Denmark
       [bf@astro.ku.dk],
       \\
       Institute for Theoretical Physics and Astrophysics,
       D-24098 Kiel,
       Germany
      }

\and

\author{Maurizio Salaris}
\affil{Astrophysics Research Institute,
       Liverpool John Moores University,
       Twelve Quays House, Egerton Wharf,
       Birkenhead L41 1LD, UK [ms@staru1.livjm.ac.uk]}

\begin{abstract}

One of the largest sources of uncertainty in the computation of globular 
cluster isochrones and hence in the age determination of globular clusters
is the lack of a rigorous description of convection.
Therefore, we calibrated the superadiabatic temperature gradient
in the envelope of metal-poor low-mass stars according to the results
from a new grid of 2D hydrodynamical models, which cover the Main Sequence and the
lower Red Giant Branch of globular cluster stars.
In practice, we still use for computing the evolutionary
stellar models the traditional mixing length formalism, but we fix the mixing
length parameter $\alpha$ in order to reproduce
the run of the entropy of the deeper adiabatic region of the stellar envelopes
with effective temperature and gravity as obtained from the hydro-models.
The detailed
behaviour of the calibrated $\alpha$ depends in a non-trivial way on the
effective temperature, gravity and metallicity of the star.
Nevertheless, the resulting isochrones for the relevant age range of 
galactic globular clusters have only small differences with respect
to isochrones computed adopting a constant solar calibrated value of the mixing
length. Accordingly, the age of globular clusters is reduced by 0.2\,Gyr at
most.

\end{abstract}


\section{Introduction}

One of the most important unsolved problems of stellar evolution
is the determination of the temperature gradient in the superadiabatic 
regions at the top of the convective envelopes of cool stars, which strongly 
affects the effective temperature ($T_{\rm eff}$) of these objects. 
The mixing length theory (MLT - B\"ohm-Vitense 1958) is widely used for
deriving this gradient. It contains a number of free parameters, among them
$\alpha$, the ratio of the mixing length to the pressure scale height, which
provides the scale length of the convection. There are different versions 
of the MLT, each one assuming different values for these parameters.
As demonstrated by Pedersen et al.~(1990), the $T_{\rm eff}$ values
obtained from the different formalisms are equivalent, provided that a suitable
value of $\alpha$ is selected (see also Gough \& Weiss 1976).
This means that the MLT results concerning stellar structure models depend only on
one free parameter, namely $\alpha$, and its absolute value depends
on the selected MLT formalism. Once the formalism is fixed,
$\alpha$ is usually calibrated by reproducing the solar $T_{\rm eff}$, and this
solar-calibrated $\alpha$ is then used for computing models of stars very
different from the Sun (e.g., metal-poor Red Giant Branch and Main Sequence
stars of various masses). However, in principle there is no compelling reason
that $\alpha$ should be the same for the Sun and different kinds of stars.

More recently, Canuto \& Mazzitelli (1991 - CM) proposed a new formalism for
the treatment of the superadiabatic convection; they take into account the full
turbulent energy spectrum and set the convective scale length equal to the
geometrical depth from the top of the convective region. Comparisons between
MLT (solar-calibrated $\alpha$) and CM stellar models
show that isochrones computed with the CM formalism cannot be
reproduced by the MLT with any constant value of $\alpha$
(Mazzitelli et al.~1995).

The problem of determining accurate effective temperatures for cool stars
affects the globular cluster (GC) age determination and, in turn, the
estimated age of the universe.
Large variations of $\alpha$ alter the derived stellar $T_{\rm eff}$ and
colours and  change the shape of GC isochrones;
as demonstrated by Chaboyer (1995 - see also Chaboyer et al.~1998)
the change of the isochrone shape can even modify the luminosity of the Turn
Off (TO - bluest point along a given isochrone). Moreover, when comparing
CM and MLT isochrones one finds that for the
most metal-poor isochrones, in the relevant range of ages of the
metal-poor galactic GCs, the TO luminosity obtained from the CM isochrones
differs appreciably from the case of the MLT (Mazzitelli et al.~1995).
Since the TO brightness is the main age indicator for GCs, uncertainties in the
convection treatment can affect the estimated GC ages (by $\approx$1\,Gyr or
even more).

In principle, one could try to constrain the convective efficiency by 
comparing theoretical isochrones with regions of
the observed Colour-Magnitude diagrams of GCs whose colours are unaffected 
by the cluster age,
but the still existing large uncertainties in the colour-transformations 
(see, e.g., Weiss \& Salaris 1998) do not permit to safely follow this approach.
An empirical constraint is given by the $T_{\rm eff}$ of the upper
Red Giant Branches of a sample of GCs as derived by Frogel et al.~(1981).
MLT models computed with a solar calibrated $\alpha$ (see, e.g., Vandenberg et
al.~1996, Salaris \& Weiss 1998) appear to be in agreement with these
empirical data, also if a precise error bar on the $\alpha$ value calibrated in
this way is hard to establish, being probably in the range $\pm$0.1-0.3 (see
Vandenberg et al.~1996 and references therein).
However, this 'empirical' calibration in principle does not constrain the
convection along the Main Sequence and the lower Red Giant Branch of GCs.

Both the MLT and CM formalisms assume a simplified, time-independent,
one-dimensional, local treatment of a typically non-stationary,
multi-dimensional, and non-local phenomenon which the stellar convection 
actually is. The
final solution to the problem of the superadiabatic convection in stellar
envelopes has to come from the computation of realistic multidimensional
radiation-hydrodynamics (RHD) simulations covering the range of effective
temperatures, gravities, and compositions typical of stars with convective
envelopes. First attempts to include in stellar models the results from rather
crude 2-dimensional (2D) and 3-dimensional (3D) hydrodynamical simulations date
back to the works by Deupree \& Varner (1980) and Lydon et al.~(1992, 1993a,b).

In this {\sl Letter} we discuss the first application to the computation of GC
isochrones of new results obtained from a grid of detailed 2D
radiation-hydrodynamics models including realistic microphysics and a
detailed treatment of the radiative transport. A further important feature of
these models is that they span a wide range in metallicity ([M/H]=$-$2.0 to
[M/H]=0.0; here we adopt the usual spectroscopic notation
[M/H]=$\log(M/H)_{\rm star}$-$\log(M/H)_{\odot}$, where M is the global metal
abundance) and cover the Main Sequence (MS) and lower Red Giant Branch (RGB)
region of GC colour-magnitude diagrams.

\section{Hydrodynamical models and calibration of the efficiency of convection}

The full grid of RHD models is described in detail elsewhere
(Ludwig et al.~1998, Freytag et al.~1998a,b), and a comprehensive discussion 
about the numerical and physical assumptions of the RHD simulations can be 
found in Ludwig et al.~(1994). Here we just recall the main
features of the models.
Each 2D model describes the atmosphere and upper layers of a star
with a convective envelope. It is obtained by solving the time-dependent,
non-linear equations of hydrodynamics for a stratified compressible fluid. The
calculations take into account a detailed treatment of the equation of state
and of the multi-dimensional, non-local, radiative transfer
(for more details see Ludwig et al.~1994). Similar to classical model
atmosphere calculations, the hydrodynamical models are fully determined by
specifying $T_{\rm eff}$, acceleration of gravity $g$, and chemical
composition, and they lie in the range
4300\,K$\le$$T_{\rm eff}$$\le$7100\,K,
2.54$\le$log($g$)$\le$4.74,
$-$2.0$\le$[M/H]$\le$0.0.

From this grid of models one can extract the entropy of the
deeper, adiabatic convective layers ($s_{\rm env}$) as a function of
$T_{\rm eff}$, $g$, and [M/H] (see Ludwig et al.~1998,
Freytag et al.~1998a,b). Once this relation is implemented in a stellar
evolution code, it completely fixes the $T_{\rm eff}$ of the star as determined
from the solution of the stellar structure equations.

A way for implementing easily this dependence of $s_{\rm env}$ on $T_{\rm eff}$
and $g$ into an evolutionary code makes use of the MLT formalism.
As explained in detail by
Ludwig et al.~(1998), for each fixed metallicity one can compute a grid of
hydrostatic one-dimensional stellar envelope models based on the MLT, covering
the same range of $g$ and $T_{\rm eff}$ spanned by the RHD computations, and
using the same input physics. By employing as surface boundary condition the
$T(\tau)$ relation derived from the hydro-models, one can calibrate an
effective $\alpha$ ($\alpha_{\rm eff}$) that is able to reproduce the
$s_{\rm env}$-$T_{\rm eff}$ relation obtained from the RHD computations.
In this way one can derive a function
$\alpha_{\rm eff}$=f($T_{\rm eff}$,$g$)
at each metallicity, that is easy to use for computing stellar
evolutionary models. The estimated error on the values of $\alpha_{\rm eff}$
derived by means of this procedure is equal to $\pm$0.05 (Ludwig et al.~1998).

Ludwig et al.~(1998) discuss the comparison of their 
RHD model for the solar envelope with the results from helioseismology.
They show that the entropy at the bottom of the superadiabatic region as
derived from helioseismology would imply a value of $\alpha_{\rm eff}$ for the
Sun slightly higher, by $\approx$0.10$\pm$0.05, than the value deduced from
their RHD models.
This small discrepancy is explained by comparing their result with the
outcome of similar 3D simulations and by the examination of the opacities used
in their models. For the Sun the 3D models predict an increase of $\alpha$ by
$\approx$0.07$\pm$0.02 with respect to the 2D ones, and the ATLAS6 opacities
(Kurucz 1979) used in the models do not consider the
contribution of the molecules. The effect of including this contribution to
the opacity would further change $\alpha$ by $\approx$0.1. The combination of 
these two effects explains the small discrepancy between the adopted solar 
RHD model and the results from helioseismology. Another important 
result derived from
the RHD models is that the effect of the envelope He abundance on the derived
$\alpha_{\rm eff}$ values is basically negligible.

The calibration of $\alpha_{\rm eff}$ for metal poor stars has been performed 
by Freytag et al.~(1998a,b), and we have used their results for computing
isochrones with typical GC metallicities [M/H]=$-$2.0 and [M/H]=$-$1.0
(scaled solar metal distribution),
Y=0.23, age $t$ ranging from 9 to 14\,Gyr, using the code
described in Salaris et al.~(1997). 
We have employed the same
$T(\tau)$ relation and the same MLT formalism (B\"ohm-Vitense 1958) used in
the calibration of $\alpha_{\rm eff}$.

We have computed a first set of isochrones using the ATLAS6 low temperature
opacities.
The only source of possible inconsistencies with the RHD models
was in this case the equation of
state (EOS) employed in the evolutionary calculations (see Salaris et al.\
1997), which is not the same as in the RHD computations.
Nevertheless, we have verified that it does 
not modify appreciably $\alpha_{\rm eff}$ as derived from the RHD calibration.
For the sake of comparison, we have computed isochrones for the same age and 
metallicities, but using a constant, solar calibrated value of $\alpha$
($\alpha_{\odot}$).
Since the ATLAS6 data do not include the contribution of the molecules to
the opacity of the stellar matter, we have repeated the same evolutionary
computations previously described (with $\alpha_{\rm eff}$ and 
$\alpha_{\odot}$), using this time the updated
Alexander \& Ferguson (1994) low temperature opacities, which include the 
molecular contribution.
When using these opacity data, we found a small effect only on the
zero point of the $\alpha_{\rm eff}$=f($T_{\rm eff}$,$g$) relation. To fix the
ideas, the solar calibration with the evolutionary code yields 
in this case $\alpha$=1.69,
while from the RHD models one gets $\alpha$=1.59 for the Sun, that means a
deviation by a factor 1.06. This small correction factor (which makes
consistent the entropy of the adiabatic layers as derived from RHD models and
from helioseismology) for the $\alpha_{\rm eff}$=f($T_{\rm eff}$,$g$) relation
has therefore been taken into account in the evolutionary calculations.

At this point we have verified that the 
differences among isochrones computed with $\alpha_{\rm eff}$ 
and $\alpha_{\odot}$ are exactly the same in the case of models computed with 
ATLAS6 or Alexander \& Ferguson (1994) opacities.
Since the latter data are a more realistic
evaluation of the opacity of stellar matter,
in the following section we will discuss the isochrones computed using
the Alexander \& Ferguson (1994) opacities.

Before concluding this section we would like to stress the
fact that the derived calibration of $\alpha_{\rm eff}$ is {\sl only intended
to reproduce the function $s_{\rm env}$($T_{\rm eff}$,$g$) of the RHD models.
The detailed temperature profile and convective velocities of the
superadiabatic layers are not represented adequately by the MLT with
$\alpha_{\rm eff}$}, but our main concern here is only the determination of
reliable effective temperatures for cool stars.

\section{Results and discussion}

Representative isochrones (9 and 13 Gyr) for the two considered metallicities,
computed with $\alpha_{\rm eff}$ and $\alpha_{\odot}$ are displayed in Figs.~1
and 2, upper panels.
The most striking feature is the close resemblance between the two sets of
isochrones. The MS loci are coincident, the $T_{\rm eff}$ values of the TO
points are very similar (notice the linear scale for the $T_{\rm eff}$ axis),
the biggest difference being equal to $\approx$45\,K for the 9\,Gyr most
metal-poor isochrone (an age possibly too young for the metal-poor galactic GC
population, see e.g.\ Salaris \& Weiss 1998). Along the RGB the isochrones
computed by using $\alpha_{\rm eff}$ are systematically hotter by only
$\approx$50\,K for [M/H]=$-$1.0 and $\approx$40\,K for [M/H]=$-$2.0.
To explain this behaviour it is useful to study the run of $\alpha_{\rm eff}$
with respect to log($L/L_{\odot}$) along the same isochrones, as shown in the 
lower panels of the same figures. 

\placefigure{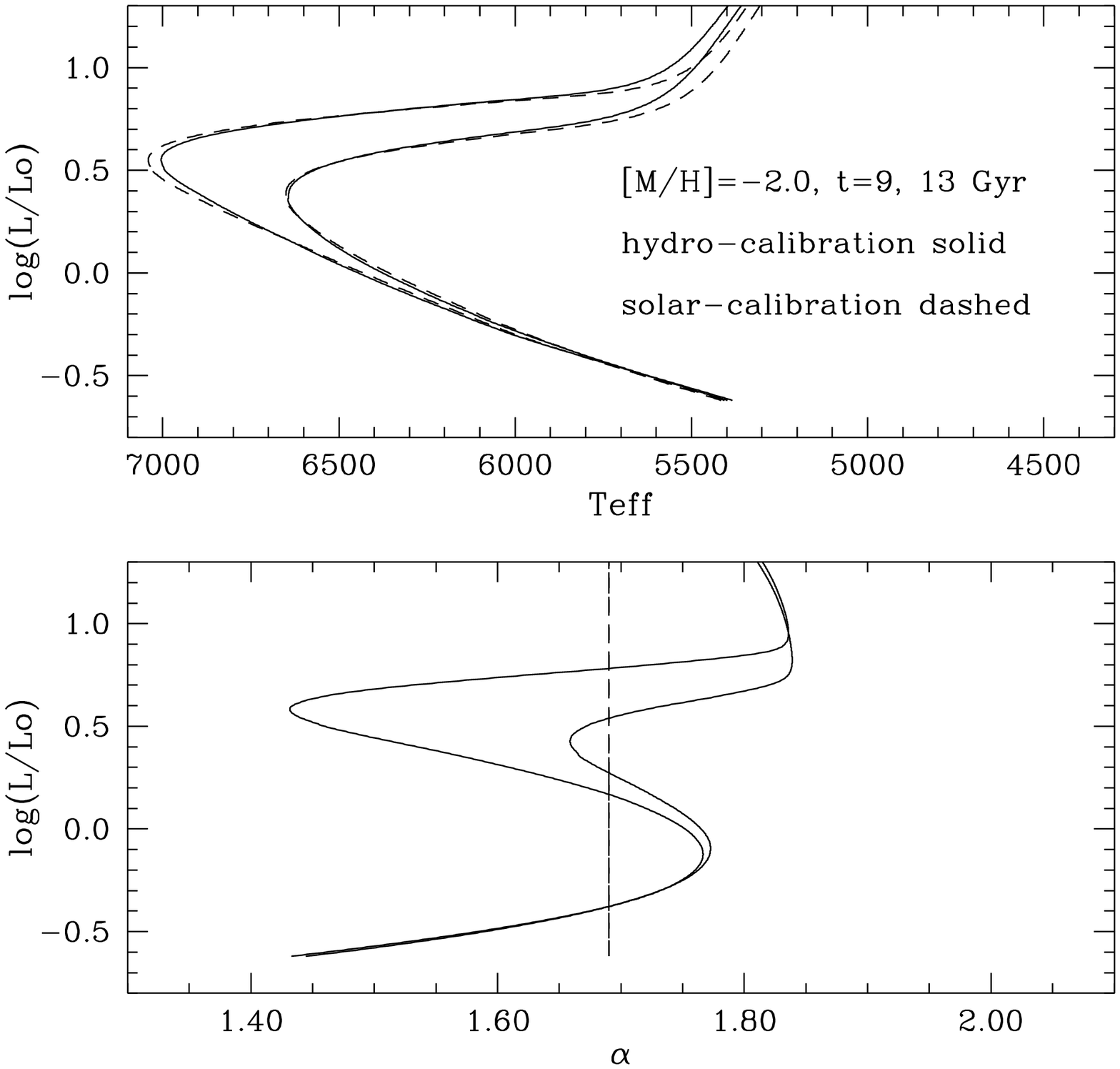}

\placefigure{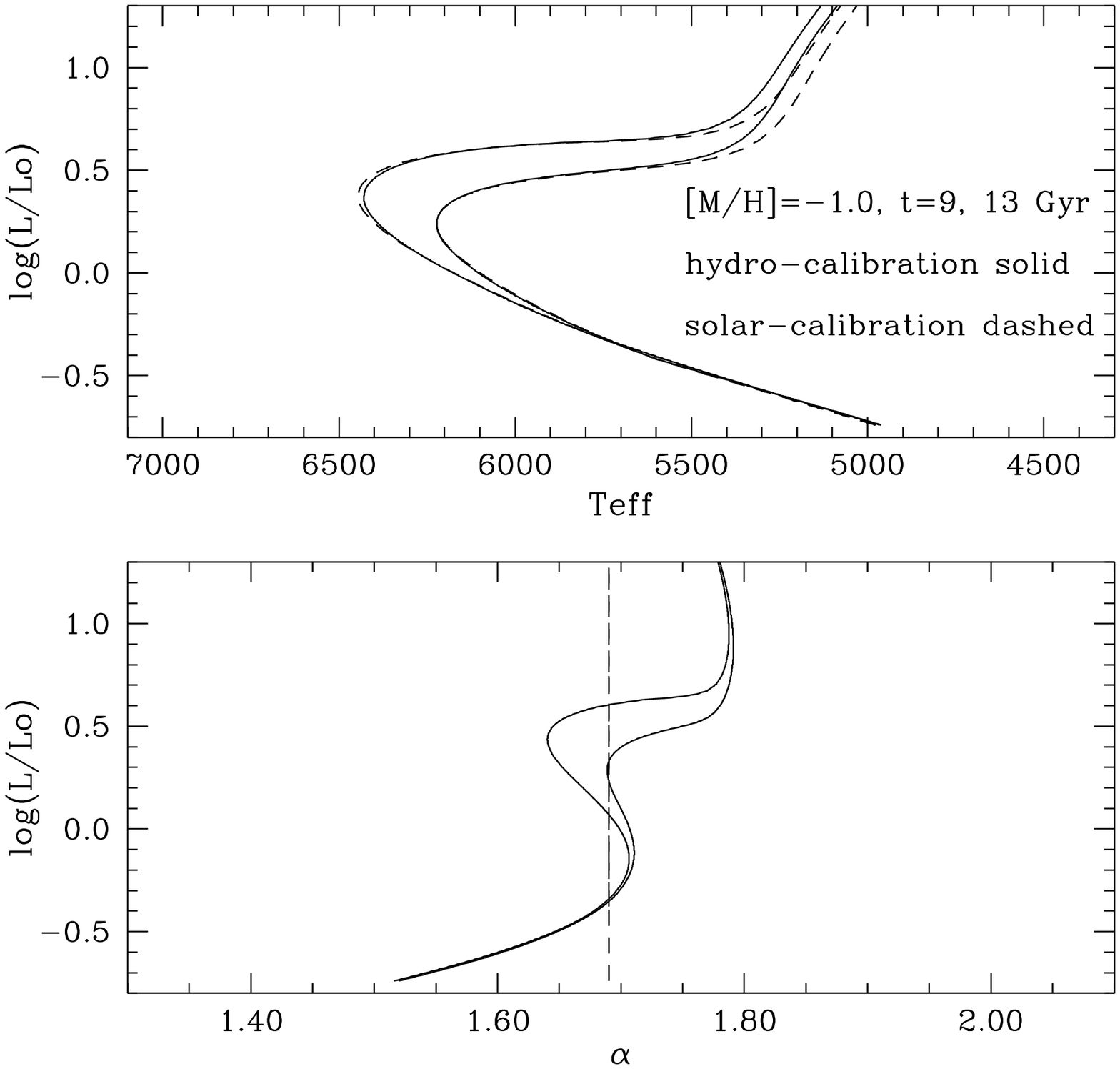}

The differences between $\alpha_{\rm eff}$ and $\alpha_{\odot}$ along the lower
MS are hardly relevant, since the $T_{\rm eff}$ of these stars is insensitive
to the choice of $\alpha$ (the entropy jump from the photosphere to the deep
adiabatically stratified layers is small anyway), while around the TO they
depend on the age of the isochrones.
In general, in the youngest, most metal-poor isochrones
$\alpha_{\rm eff}$ shows the largest difference with respect to
$\alpha_{\odot}$, but since stars in these phases are quite hot ($T_{\rm
eff}$$\approx$7000\,K) and their convection zones are relatively shallow,
the sensitivity of $T_{\rm eff}$ to $\alpha$ is not very large.
Along the RGB, where the $T_{\rm eff}$ of stellar models is most sensitive to
$\alpha$ because of deeper superadiabatic regions, 
$\alpha_{\rm eff}$ is systematically higher than $\alpha_{\odot}$
by 0.10-0.15 for both metallicities. This difference causes a systematic 
shift by $\approx$50\,K toward higher $T_{\rm eff}$ with respect to the case 
of $\alpha_{\odot}$, a quantity marginally significant 
since the error by $\pm$0.05 on $\alpha_{\rm eff}$ translates into an error by 
$\approx$$\pm$15-20\,K on the RGB $T_{\rm eff}$).
Qualitatively, the behaviour of isochrones computed using $\alpha_{\rm eff}$
looks similar, for certain features, to the results of the CM formalism;
we are referring here to the fact that the TO is cooler (but only
for the youngest more metal-poor isochrones) than for the models computed with
$\alpha_{\odot}$. But the differences we find are smaller than the
predictions of the CM formalism. Moreover, the RGB location in the 
$\alpha_{\rm eff}$ isochrones is only slightly hotter than in the $\alpha_{\odot}$
ones, while in the case of CM models the RGB is cooler at low metallicity and 
progressively hotter for increasing metallicities.

At this point, let's turn our attention to the GC age indicators that one can
extract from the isochrones. The TO brightness is the most solid one; once the
distance is fixed (e.g., from the Horizontal Branch luminosity or by means of
the subdwarf fitting technique), the comparison between theoretical and
observed TO gives directly the cluster age. The TO colour is also a possible
age indicator once the reddening is known, but the present
uncertainties on the colour transformations do not favour this method for
deriving absolute ages, also if the isochrone $T_{\rm eff}$ and the GC
reddenings are determined with high accuracy. A third possibility is to use
the reddening- and distance modulus-independent quantity $\Delta(B-V)$, that is
the difference in (B-V) between TO and base of the RGB, as defined by
Vandenberg et al.~(1990). Again, the uncertainties on the
colour-transformations prevents from using the absolute value of $\Delta(B-V)$
for deriving absolute GC ages, but the differential use of this quantity is a
solid and widely employed indicator of age differences (see, e.g., Vandenberg
et al.~1990, Salaris \& Weiss~1997), and it is weakly dependent on [M/H].

In Figure~3 (upper and center panel) we compare the TO position (brightness and
colour) in the age range 9-14\,Gyr for the two sets of isochrones with
$\alpha_{\rm eff}$ and $\alpha_{\odot}$. We have transformed the isochrones to
the observational V-(B-V) plane according to the 
colours and bolometric corrections used by Salaris \& Weiss (1998), 
but the results of
this comparison do not depend on the particular set of transformations used.
As it is evident from the figure, the age differences as derived from the TO
brightness (or colour) are basically negligible.

\placefigure{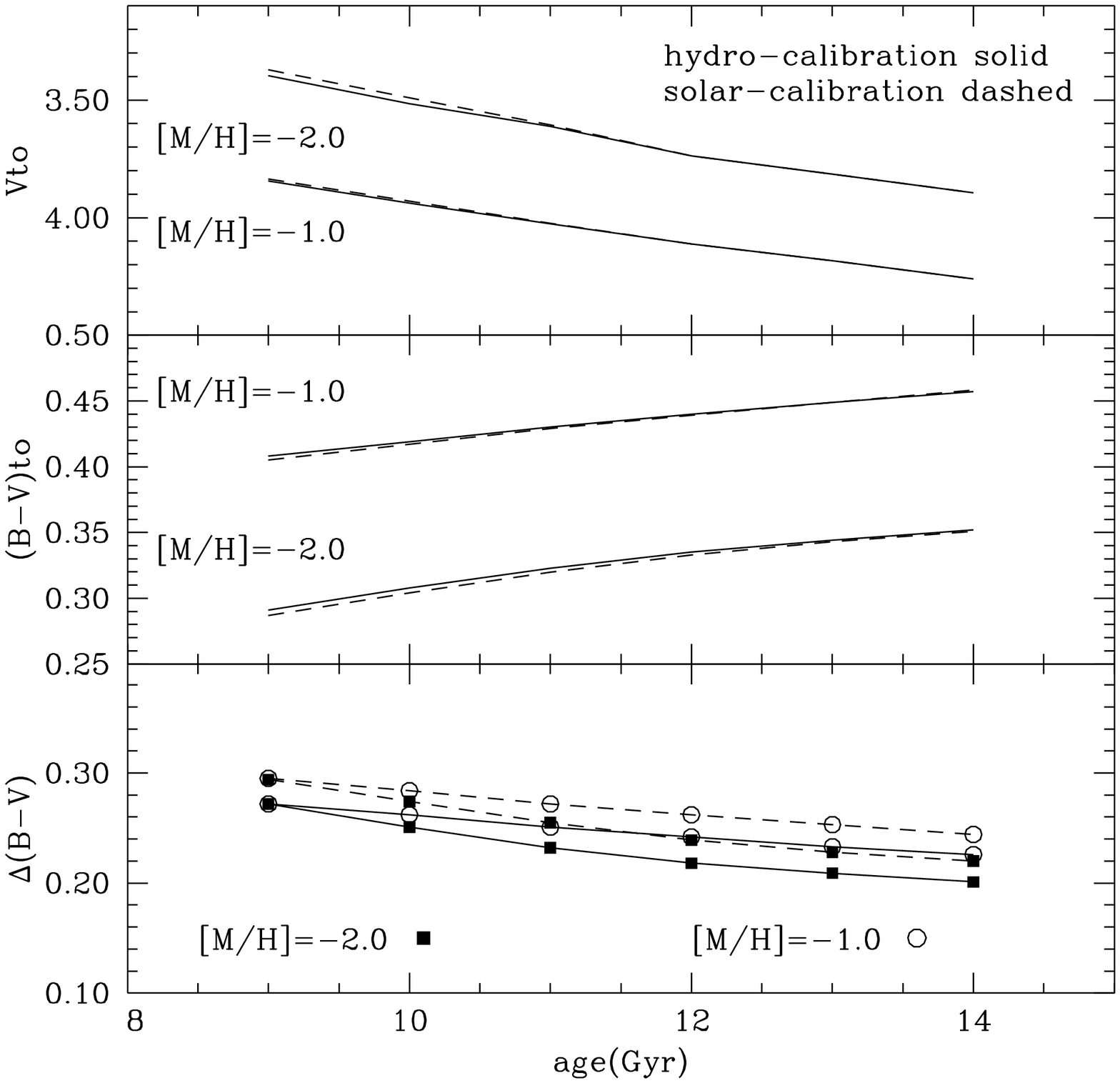}

In the lowest panel of Figure~3 we compare $\Delta(B-V)$ as a
function of the age, for both sets of isochrones. At each metallicity the two
curves corresponding to the two calibrations of the convection lie parallel for
all the relevant age range (the absolute values being different by only
$\approx$0.02\,mag). Therefore, the derivative $\delta (\Delta(B-V))/\delta t$
(and the relative ages derived for the $\Delta(B-V)$) is not affected at all
when $\alpha_{\rm eff}$ is used instead of $\alpha_{\odot}$.

In conclusion, the main results of this analysis show that the $T_{\rm eff}$ of
GC isochrones computed employing the MLT formalism and $\alpha_{\odot}$, or the
$\alpha_{\rm eff}$ calibration as derived from detailed RHD models are in good
mutual agreement: the maximum deviations in the relevant age range for galactic
GC amounts at most to a systematic shift by $\approx$~50$\pm$20\,K
along the RGB. As previously discussed, preliminary comparisons 
(Ludwig et al.~1998) between the adopted grid of 2D RHD models and a small 
sample of 3D ones show only a very small systematic shift of $\alpha_{\rm eff}$
as derived from the RHD
models by $\approx$0.07, which does not affect our results appreciably.

The next necessary step for finally solving the problem of superadiabatic
convection in stellar envelopes involves the computations of 3D model grids
with up-to-date equation of state and frequence-dependent opacity tables
to improve especially the photospheric temperature structure.
Particularly the resolution of the numerical grid in the vertical direction 
should be improved to resolve the extremely sharp temperature jump at the 
bottom of the photosphere when the computations are extended to higher 
luminosities. A better coverage of the transition region from efficient to
weak (radiation dominated) convection at high effective temperatures would
improve the base to judge between the MLT and the CM formalism.
Nevertheless, the already existing grid of 2D models indicates that in the
$T_{\rm eff}$-$\log g$-[M/H] region of interest for GC stars
there is no drastic change in the properties of the envelope convection.
Accordingly, the use of the MLT with a constant solar-calibrated
$\alpha_{\odot}$ leads only to insignificant errors of at most 0.2\,Gyr in the
derived ages of globular clusters. 

It is important also to remark again that
the structure of the superadiabatic convective regions is not 
suitably reproduced either by $\alpha_{\odot}$ nor by $\alpha_{\rm eff}$,
and that the complete results from RHD models have to be employed whenever a
detailed description of the properties of this layers is needed
(e.g., for astro- and helioseismology).
In addition, the RHD models should be analysed
regarding the effects on the $T_{\rm eff}$-colour-transformations.

\acknowledgments

We would like to express our gratitude to Hartmut Holweger and Matthias Steffen
for continuous support and helpful discussions. We warmly thank John Porter,
Achim Weiss, and Hans-G\"unter Ludwig for reading an early draft 
of the manuscript and for useful suggestions. B.F.\ was supported by the
{\em Deutsche Forschungsgemeinschaft}, DFG project Ho~596/36-1 and by the
European {\em Training and Mobility of Researchers\/} Programme.

% And finally, we must deal with the figures.  There are three figures
% associated with this manuscript; two figures are Encapsulated
% PostScript (EPS) files.  The third figure is a grey scale figure that does
% not exist in EPS form.
%
% Authors have three options for including figure information within a 
% manuscript.  Not all the options may be acceptable by the target Journal - be
% sure to look at the appropriate submission instructions, electronic or 
% otherwise.
%
% Option 1.  Using this option, only the figure captions are included in the
% main body of the manuscript.  The figure captions must start on a new page.
% The captions are generated with the \figcaption[]{} command: the first 
% argument is optional, if you put something in there, put the name of the 
% EPS file that goes with the caption; the second argument is the figure 
% caption itself, and may include a \label command.  The \figcaption command
% generates the figure numbers.  This option is acceptable for all manuscript
% submissions.

\clearpage

\figcaption[fig1pap.eps]{Upper panel: isochrones in the 
$T_{\rm eff}$-log($L/L_{\odot}$)
diagram for [M/H]=$-$2.0. 
Two isochrones (t=9, 13 Gyr) computed with $\alpha_{\rm eff}$ are compared
to the corresponding isochrones computed with a constant
$\alpha_{\odot}$. Lower panel: values of $\alpha_{\rm eff}$ along the 
hydro-calibrated isochrones; the vertical line represents the 
constant $\alpha_{\odot}$ of the standard isochrones.\label{fig1}}

\figcaption[fig2pap.eps]{As in Figure 1 but for [M/H]=$-$1.0.\label{fig2}}

\figcaption[fig3pap.eps]{Upper panel: $M_{v}$ values for the TO of 
isochrones computed by using either $\alpha_{\rm eff}$ or $\alpha_{\odot}$, 
[M/H]=$-$2.0 and $-$1.0, t=9-14\,Gyr. Central panel: the same, but for 
the TO (B-V) colour. Lower panel: the same, but for the quantity 
$\Delta(B-V)$.\label{fig3}}

%\end{document}

% Option 2.  The figure captions are printed on a caption page(s) as in 
% option 1.  The figures available as EPS files are then printed at the
% end of the document, one figure per page, using the \plotone command.
% If you wish to process this option then simply comment out the \end{document}
% just above these five lines. 

\clearpage

\plotone{fig1pap.eps}

\clearpage

\plotone{fig2pap.eps}

\clearpage

\plotone{fig3pap.eps}

\clearpage

\end{document}